\documentclass[11pt]{article}

\usepackage{graphicx}
\usepackage[bookmarksnumbered,colorlinks,plainpages]{hyperref}

\begin{document}

\title{Generalized algebra within a nonextensive statistics}

\author{L. Nivanen, A. Le M\'ehaut\'e and Q.A. Wang
\\ Institut Sup\'erieur des Mat\'eriaux du Mans, \\
44, Avenue F.A. Bartholdi, 72000 Le Mans, France \\ Email: awang@ismans.univ-lemans.fr}

\date{}

\maketitle

\begin{abstract}
By considering generalized logarithm and exponential functions used in nonextensive
statistics, the four usual algebraic operators : addition, subtraction, product and division,
are generalized. The properties of the generalized operators are investigated. Some standard
properties are preserved, e.g., associativity, commutativity and existence of neutral
elements. On the contrary, the distributivity law and the opposite element is no more
universal within the generalized algebra.
\end{abstract}



\section{Introduction}
Although scientists apply Boltzmann-Gibbs statistics (BGS), or its exponential factor of
probability distribution, to systems having long range correlation or finite
size\cite{Conform}, this classical statistical theory, from the usual point of view, remains
an additive theory in the thermodynamic limits, i.e., the extensive thermodynamic quantities
are proportional to its volume or to the number of its elements.

As far as we know, there is no direct experimental evidence of nonadditive energy. However,
theoretical and numerical work has disclosed many models characterized by nonadditive energy
or entropy with, e.g., black hole\cite{Hayw98,Frol96} and some magnetic and fluid models
including long range interaction\cite{Jund95,Cann96,Ruff95,Grig96,Ruff01}. Interesting
numerical results also showed that, in the case of above long range (nonintegrable)
interactions, in the thermodynamic limits, the systems show complex behaviors at the edge of
chaos with non-gaussian distribution, anomalous diffusion and dynamic correlation in the phase
space\cite{Lato02}. For this kind of non-equilibrium systems at stationary
state\cite{Letters}, a possible way out has been suggested on the basis of a nonadditive
statistical mechanics (NSM)\cite{Tsal88,Cura91,Penni,Tsal99,Wang01,Wang02,Wang02b,Wang02c}.

NSM introduces\cite{Tsal88,Tsal94} generalized distribution functions called $q$-exponential
given by $exp_q(x)=[1+(1-q)x]^{1/(1-q)}$ or $exp_q^*(x)=[1+(q-1)x]^{1/(q-1)}$ according to
different formalisms related to incomplete information\cite{Wang01,Wang02} and complete
information hypothesis\cite{Tsal88,Cura91,Penni,Tsal99}. The above functions are the inverse
functions of the generalized logarithm $\ln_q(x)=\frac{x^{1-q}-1}{1-q}$ or
$\ln_q^*(x)=\frac{x^{q-1}-1}{q-1}$ which can be used as a generalization of Hartley
logarithmic information measure\cite{Wang01}. In NSM, we require that the $[.]$ in
$q$-exponential be positive to assure physically significant probability distribution. This
requirement may lead to high or low energy cutoff depending on the value of $q$. In what
follows, we will write the generalized functions as follows : $\ln_a(x)=\frac{x^a-1}{a}$ and
$e_a^x=exp_a(x)=[1+ax]^{1/a}$. On the one hand, these forms are simpler to treat
mathematically. On the other hand, this choice ensures the generality of the present results
which should be independent of the physical circumstances with different formalisms of NSM.
$e_a(x)$ and $\ln_a(x)$ tend to the conventional counterparts whenever $a\rightarrow 0$.

Although defined originally from the physical viewpoint, these functions present important
mathematical interests. Some of their important properties have been studied recently in
\cite{Yama02,Agui03}. They have also been used to generalize hyperbolic function and
algebra\cite{Borges}. In this paper, we discuss several properties of the new algebra
generated by the $q$-logarithm and $q$-exponential. This algebra can be viewed as a
generalization of the conventional additive algebra related to the normal logarithm and
exponential, but characterized by a nonadditive additional factor. A similar generalization of
the conventional algebra has been proposed by Kaniadakis on the basis of the $\kappa$-deformed
logarithm and exponential\cite{Kaniadakis}.

The paper is organized as follows. In the following section we briefly discuss the
essential aspects of NSM. The generalized algebra is discussed in the third section.

\section{Some nonadditive relations of NSM}
We know that, for a nonadditive thermodynamic system, the existence of thermodynamic
stationarity may be used as a constraint upon the form of the nonextensivity of physical
quantities\cite{Abe01}. For the entropy $S$ and the internal energy $U$, this can be discussed
rigorously by using the conventional BGS method for the statistical interpretation of the
zeroth law. It has been shown\cite{Abe01} that the simplest nonextensivities prescribed by the
thermodynamic stationarity condition were generated by the following relationships :
$S(A)=\frac{h(A)-1}{\lambda_S}$, $S(B)=\frac{h(B)-1}{\lambda_S}$,
$S(A+B)=\frac{h(A+B)-1}{\lambda_S}$ and $h(A+B)=h(A)h(B)$, where $A$ and $B$ are two
subsystems of an isolated system $A+B$, $\lambda_S$ is a constant, $h(A)$ or $h(B)$ is the
factor depending on $A$ or $B$ in the derivative $\frac{\partial S(A+B)}{\partial S(B)}$ or
$\frac{\partial S(A+B)}{\partial S(A)}$. It was also found that these relationships are also
valid if $S$ is replaced by $U$ and $\lambda_S$ by $\lambda_U$\cite{Wang02d}. So the
nonextensivities should be the following : $S(A+B)=S(A)+S(B)+\lambda_SS(A)S(B)$ and
$U(A+B)=U(A)+U(B)+\lambda_UU(A)U(B)$. We have shown\cite{Wang02} that only these two
relationships allow to interpret the zeroth law of thermodynamics within NSM and define a
physical temperature without any approximation.

The above nonextensivity of entropy, according to some authors\cite{Santos}, uniquely
leads to the entropy
\begin{equation}\label{1}
S=-k\frac{\sum_ip_i-\sum_ip_i^q}{1-q},
\end{equation}
which can be shown\cite{Wang01,Wang02} to be the following expectation value
$S=\sum_ip_i^qI_i$ of a generalized information measure :
\begin{equation}\label{1a}
I_i=k\ln_{1-q}(p_i)
\end{equation}
where $q=1+\lambda_S$ is a real parameter\cite{Tsal88}, $k$ is Boltzmann constant and $p_i$
the probability that the system is found at the state $i$, if it is supposed that
$p_{ij}(A+B)=p_i(A)p_j(B)$. This product law of probability has always been considered as a
result of the ``statistical independence'' of $A$ and $B$ and considered by many as a
``reason'' for accepting additive energy $U(A+B)=U(A)+U(B)$ within NSM. As shown with the
counter-example, this is not necessarily true and entails several fundamental and practical
problems which have been discussed in detail\cite{Wang02}. We observed\cite{Wang02e} that the
product law of probability could be considered as a consequence of the entropy in Eq.(\ref{1})
and the nonextensivity imposed by the condition of the existence of equilibrium. In this way,
the composite energy is freed from the constraint of the independence of the subsystems and
NSM is then entitled to treat nonadditive systems having the nonadditive entropy postulated in
Eq.(\ref{1}) (or the information measure postulated in Eq.(\ref{1a}) and the nonadditive
energy prescribed by the thermal equilibrium.

The maximization of the above entropy leads to the $q$-exponential probability distribution.
The currently improved coherence of this statistics and the investigations of the
relationships between this statistics and the chaotic or fractal phenomena\cite{Wang03} tell
us that $q$-logarithm and $q$-exponential have solid physical background and may play a major
role in the nonadditive physics. So it is worthwhile to investigate them from pure
mathematical point of view and especially according to the related algebra associated with
nonadditive properties. In what follows, we present the generalized nonadditive algebra
related to these functions.

\section{Generalized operations}
From now on, the mathematical notion of morphism will be used to exhibit generalized
expressions for the four fundamental algebraic operators : addition, subtraction, product
and division. In the fourth part, the reader will see that the algebraic properties of
these generalized operators are not always identical to those of the classical operators.
New features will be pointed out, for instance, the distributive law is no more verified.

The standard exponential and logarithm functions present some remarkable properties. For
instance, the exponential function is a morphism from $(\mathbf{R}, +)$ to
$(\mathbf{R}_+^*,\times)$. Through reciprocity, the logarithm function is a morphism from
$(\mathbf{R}_+^*,\times)$ to $(\mathbf{R}, +)$. As a result : $e^{x+y}=e^x e^y$ and $\ln
xy = \ln x+\ln y$. Although these properties cannot be generalized directly by a simple
substitution of $e^x$ by $e_a^x$ (or $\ln x$ by $\ln_a x$, the following slightly more
complicated relationships are verified :

\begin{equation}\label{2}
e_a^x e_a^y=e_a^{x+y+axy}\neq e_a^{x+y},
\end{equation}

\begin{equation}\label{3}
e_a^{x+y}=[(e_a^x)^a+(e_a^y)^a-1]^{1/a}\neq e_a^x e_a^y,
\end{equation}

\begin{equation}\label{4}
\ln_a x +\ln_a y=\ln_a (x^a+y^a-1)^{1/a}\neq \ln_a xy,
\end{equation}

\begin{equation}\label{5}
\ln_a xy = \ln_a x + \ln_a y + a\ln_a x \ln_a y\neq \ln_a x + \ln_a y.
\end{equation}

We see that with standard addition and product operators, the concept of morphism cannot
be applied to generalized functions. But if we $define$ generalized addition (denoted by
$+_a$) and product ($\times_a$) operators that depend on the parameter $a$ as follows :

\begin{equation}\label{a}
x+_ay = x+y+axy
\end{equation}
and
\begin{equation}\label{b}
x\times_ay = (x^a+y^a-1)^{1/a},
\end{equation}
then Eqs.(\ref{2}) to (\ref{5}) can be recast into :

\begin{equation}\label{2a}
e_a^{x+_ay}=e_a^x e_a^y,
\end{equation}

\begin{equation}\label{3a}
e_a^{x+y}=e_a^x \times_a e_a^y,
\end{equation}

\begin{equation}\label{4a}
\ln_a x\times_ay=\ln_a x +\ln_a y,
\end{equation}

\begin{equation}\label{5a}
\ln_a xy = \ln_a x +_a \ln_a y .
\end{equation}

The two standard morphisms have split into four ones. Let us denote the definition set of
the function $e_a^x$ by $D_a$. From Eq.(\ref{2a}), it comes that the generalized
exponential function is a morphism from $({D_a}, {+_a})$ to $(\mathbf{R}_+^*,\times)$.
But from Eq.(\ref{3a}), we note that this function is also a morphism from $({D_a}, +)$
to $({\mathbf{R}}_+^*,{{\times}_a})$. From Eq.(\ref{4a}), it comes that the generalized
logarithm function is a morphism from $({\mathbf{R}}_+^*,{{\times }_a})$ to $({D_a}, +)$.
But from Eq.(\ref{5a}), we note that this function is also a morphism from
$(\mathbf{R}_+^*,\times)$ to $({D_a}, {+_a})$. It is worth noticing that a standard
operator and a generalized one are present in each of the Eqs.(\ref{2a}) to (\ref{5a}).
This is because that the generalized operators are defined with the help of the standard
ones in Eqs.(\ref{a}) and (\ref{b}).

In the same way, the generalized subtraction $-_a$ and division $/_a$ operators can be
defined. If we write :

\begin{equation}\label{aa}
x-_ay = \frac{x-y}{1+ay}
\end{equation}
and
\begin{equation}\label{bb}
x/_ay = (x^a-y^a+1)^{1/a}
\end{equation}
then the following relationships exist :

\begin{equation}\label{2b}
e_a^{x-_ay} = e_a^x /e_a^y
\end{equation}

\begin{equation}\label{3b}
e_a^{x-y}=e_a^x /_a e_a^y
\end{equation}

\begin{equation}\label{4b}
\ln_a(x /_a y) = \ln_a x - \ln_a y
\end{equation}

\begin{equation}\label{5b}
\ln_a(x/y)=\ln_a (x) {-_a} \ln_a y.
\end{equation}
which recover the standard ones $e^{x-y}=e^x/e^y$ and $\ln(x/y)=\ln x -\ln y$ when
$a\rightarrow 0$.

\section{Properties of generalized operators}

\subsection{Additivity}
The generalized addition operator has following properties :

\begin{enumerate}

\item Associativity : $(x+_ay)+_az=x+_a(y+_az)$.

\item Commutativity : $x+_ay=y+_ax$.

\item 0 is the neutral element, i.e., $x+_a0=0+_ax=x$.

\item If $x\neq x_0=-1/a$, it has an opposite element $-_ax$, i.e., $-_ax = -x/(1 + ax)$.
The fact that $x_0$ has no opposite element is a new feature. It should be noticed that
the above definition of opposite element is compatible with the definition of
substraction : $x {-_a} y=x +_a (-_ay)$.

\item The sign rules : ${-_a}({-_a}x)={+_a}({+_a}x)=x $ and
${-_a}({+_a}x)={+_a}({-_a}x)={-_a}x$.

\item Generating role of $+_1$ : if we note $Z=x+_ay$, then $aZ=(ax)+_1(ay)$.

\end{enumerate}

\subsection{Multiplication}
The product operator $\times_a$ has more complicated features. For a given $a$, we see
from Eq.(\ref{4}) that, if $x,y > 0$, the product $x\times_ay$ is defined unambiguously
and the inequality ${x^a} + {y^a} > 1$ is verified. We notice following properties :

\begin{enumerate}

\item Associativity : $(x\times_ay)\times_az=x\times_a(y\times_az)$.

\item Commutativity : $x\times_ay=y\times_ax$.

\item 1 is the neutral element, i.e., $x\times_a 1=1\times_a x=x$.

\item There is no absorbing element (like zero in the usual case). We cannot find a real
number $y$ such that, for arbitrary real number $x$, we have $x \times_a y = y$.

\item x has an inverse element noted $1{/_a} x = (2 - x^a)^{1/a}$. In particular, we see that
if $a > 0$, then $1{/_a} 0 = 2^{1/a}$ and $1/_a 2^{1/a} = 0$. So 0 can have a finite inverse
in the generalized algebra.

\item Generating role of $\times_1$ : if we note $Z=x\times_ay$, then
$Z^a=(x^a)\times_1(y^a)$.
\end{enumerate}

We notice that the new operators of this generalized algebra have more complex properties
than the usual ones. Unlike the standard case, the existence of an opposite element is
not automatic. On the other hand, 0 can now be inverted. Now with the new operators, the
infinity concept may be taken into account with finite numbers. There is a strong analogy
between this property and the role of hyperbolic space in metric topology\cite{Bear}. It
is natural to find these two distinct mathematical tools in the study of physical
phenomena taking place on complex media characterized by a set of singularities in a
compact space\cite{Leme1,Leme2}.

Another important feature of the new algebra with the operators defined previously is the
disparition of the law of distributivity, i.e., $x y +x z = x (y + z)$ within the usual
algebra. With the above generalized operators, no combination is possible, except for a
particular case like x = 1. It is straightforward to show that, {\it in general}, we have :

\begin{equation}\label{6}
x {{\times }_a} y + x {{\times }_a} z \neq x {{\times }_a} (y + z)
\end{equation}

\begin{equation}\label{7}
x y {+_a} x z \neq x (y {+_a} z)
\end{equation}

\begin{equation}\label{8}
x {{\times }_a} y {+_a} x {{\times }_a} z \neq  x {{\times }_a} (y {+_a} z)
\end{equation}
This non distributivity has important consequences for the manipulation of analytical
expressions. The most important one is the impossibility to develop or factorize expressions.

On the other hand, distributivity can be recovered by defining different generalized addition
and product operators as has been done in \cite{Kaniadakis} :

\begin{equation}                                \label{9}
\ln[e_a^{x{\times^a}y}] = \ln e_a^x\ln e_a^y
\end{equation}
and
\begin{equation}                                \label{10}
e^{[\ln_a[x{+^a}y]} = e^{\ln_ax} + e^{\ln_a y}.
\end{equation}
The distributivity can be established with either $\times^a$ and $+_a$ or $\times_a$ and
$+^a$, i.e. :
\begin{equation}                                \label{11}
x {\times^a} y +_a x{\times^a}z = x {{\times }^a} (y +_ a z)
\end{equation}
or
\begin{equation}                                    \label{12}
x {\times_a} y +^a x{\times_a}z = x {\times_ a} (y +^a z).
\end{equation}
It should be noted that the distributivity does not exist neither between the $\times^a$ and
$+ ^a$ defined in Eqs.(\ref{9}) to (\ref{10}), i.e. :
\begin{equation}                                \label{13}
x {\times^a} y +^a x{\times^a}z \neq x {\times^a} (y +^a z),
\end{equation}
nor between these two operators and the ordinary operators.

The operators $\times^a$ and $+^a$ are respectively given by
\begin{equation}                                \label{14}
x {\times^a} y = \frac{e^{\ln(1+ax)\ln(1+ay)/a}-1}{a}
\end{equation}
and
\begin{equation}                                \label{15}
x {+^a} y = [a\ln(e^{x^a/a}+e^{y^a/a})]^{1/a},
\end{equation}
which are different from $\times_a$ and $+_a$ and have different properties. For example, we
have here $x {+^a}0\neq x$, $x {\times^a} 1\neq x$ and $x {\times^a} 0 = 0$.

\section{Examples of simple application}
It is obvious that, if we use this generalized algebra with the operators defined in
Eqs.(\ref{a}) to (\ref{5a}), the nonadditive statistical mechanics and thermodynamics will be
able to be expressed in additive form just as in BGS. In one of our papers\cite{Wang02c}, NSM
is given in additive form by deforming the nonadditive energy and entropy. The same formalism
can be given with generalized algebra without deforming the nonadditive physical quantities.
That is, for a total system composed of two subsystems $A$ and $B$, we can write for entropy
$S(A+B)=S(A)+_{a_S}S(B)$ and for energy $U(A+B)=U(A)+_{a_U}U(B)$ where
$a_S=-\lambda_S=(1-q)/k$ and $a_U=-\lambda_U=(1-q)/kT$ for NSM.

Let us consider the temporal evolution of the price of a product at successive 1st of
January. It costs respectively $x_0 = 100\$$ in 2001, ${x_1} = 110\$$ in 2002 and ${x_2}=
121\$$ in 2003. The annual evolution ratio is ${y_0} = ({x_1} - {x_0} )/{x_0} = 0,1$ and
${y_1} = ({x_2} - {x_1} )/{x_1} = 0.1$, or $10\%$ each time. The global evolution ratio
over two years is $y'_0 = ({x_2} - {x_0} )/{x_0} = 0.21$, that is $21\%$. Of course
$y'_0\neq y_0+y_1$ because the ratios are not based upon the same denominator. However,
simple calculation leads to $y'_0=y_0+y_1+y_0 y_1$. This expression is rewritten with the
help of the generalized addition operator defined in Eq.(\ref{a}) : $y'_0 = y_0 +_1 y_1$.
It means that combinations of ratios can be considered as additive by using the
generalized addition. In this case, $a=1$ is a universal value. This can be shown for
whatever ratios as follows. From Eq.(\ref{a}), we have
$a=(\frac{x_2-x_0}{x_0}-\frac{x_1-x_0}{x_0}-\frac{x_2-x_1}{x_1})/\frac{(x_1-x_0)(x_2-x_1)}{x_0x_1}
=\frac{x_1-x_0}{x_1-x_0}=1$. This calculation can be useful for other temporal process.

\section{Conclusion}

In this paper some algebraic aspects of the nonadditive statistics have been studied. The four
classical operators of the usual algebra have been generalized. The new operators allow to
preserve the morphism properties of the exponential and logarithm functions. Some properties
(associativity, commutativity, existence of neutral element and the sign rules) of the
standard operators can be extended to the generalized ones. But others cannot, e.g. the
opposite element does not exist for arbitrary element; and the generalized additions and
multiplications are not always distributive. Another interesting point should be noted : in
the generalized formalism, ``0" is no more the absorbing element of multiplication but can be
inverted just like other element.

We hope that this generalized algebra can be helpful for the understanding and the development
of nonadditive physics which seems inevitable in view of the complex phenomena that cannot be
described within the additive statistical theory.


\begin{thebibliography}{99}

\bibitem {Conform}
J.L. Cardy, Conformal invariance and statistical mechanics, in {\it Fields, strings and
critical phenomena, Les Houches 1988}, Ed. by E. Brézin and J. Zinn-Justin, North Horland
(1990)p168

B. Jancovici, G. Manificat and C. Pisani, Coulomb systems seen as critical systems :
finite-size effects in two dimensions, {\em J. Stat. Phys.,\/} {\bf 70}(1994)3147

\bibitem {Hayw98}
Sean A. Hayward, {\em Class. Quant. Grav.,\/} {\bf 15}(1998)3147, gr-qc/9710089

Sean A. Hayward, {\em Phys. Lett. A,\/} {\bf 256}(1999)347-350

\bibitem {Frol96}
V.P. Frolov, D.V. Fursaev, and A.I. Zelnikov, {\em Phys. Lett. B,\/} {\bf 382}(1996)220,
Hep-th/9603175

\bibitem {Jund95}
P. Jund, S.G. Kim and C. Tsallis, {\em Phys. Rev. B,\/} {\bf 52}(1995)50

\bibitem {Cann96}
S.A. Cannas and F.A. Tamarit, {\em Phys. Rev. B,\/} {\bf 54}(1996)R12661

\bibitem {Ruff95}
M. Antoni and S. Ruffo, {\em Phys. Rev. E,\/} {\bf 52}(1995)2361

\bibitem {Grig96}
J.R. Grigera, {\em Phys. Lett. A,\/} {\bf 217}(1996)47

\bibitem {Ruff01}
J. Barr\'e, D. Mukamel and S. Ruffo, {\em Phys. Rev. Lett.,\/} {\bf 87}(2001)030601

J. Barr\'e, D. Mukamel and S. Ruffo, {\em Lecture Notes in Physics,\/} {\bf
602}(2002)45-67

T. Dauxois, S. Ruffo, E. Arimondo, and M. Wilkens, {\em Lecture Notes in Physics,\/} {\bf
602}(2002)1-19

\bibitem {Lato02}
V. Latora, A. Rapisarda, C. Tsallis, {\em Physica A,\/} {\bf 305}(2002)129

V. Latora, A. Rapisarda, C. Tsallis, {\em Phys. Rev. E,\/} {\bf 64}(2001)056134

V. Latora, A. Rapisarda, S. Ruffo, {\em Prog. Theor. Phys. Suppl.,\/} {\bf 139}(2000)204

\bibitem {Letters}
S. Abe, A.K. Rajagopal, A. Plastinos, V. Lotora, A. Rapisarda and A. Robledo, Letters to the
Editor : Revisiting disorder and Tsallis statistics, {\em Science,\/} {\bf 300}(2003)249-251

\bibitem {Tsal88}
C. Tsallis, {\em J. Stat. Phys.,\/} {\bf 52}(1988)479

\bibitem {Cura91}
EMF. Curado and C. Tsallis, {\em J. Phys.A:Math.Gen.,\/} {\bf 24} (1991)L69

\bibitem {Penni}
F. Pennini, A.R. Plastino and A. Plastino, {\em Physica A,\/} {\bf 258}(1998)446

\bibitem {Tsal99}
C. Tsallis, R.S. Mendes and A.R. Plastino, {\em Physica A,\/} {\bf 261}(1999)534;

Silvio R.A. Salinas and C. Tsallis, {\em Brazilian Journal of Physics(special issue:
Nonadditive Statistical Mechanics and Thermodynamics),\/} {\bf 29}(1999).

\bibitem {Wang01}
Q.A. Wang, Incomplete statistics : nonextensive generalization of statistical mechanics, {\em
Chaos, Solitons $\&$ Fractals,\/} {\bf12}(2001)1431

\bibitem {Wang02}
Q.A. Wang, Nonextensive statistics and incomplete information, {\em Euro. Phys. J. B,}
{\bf26}(2002)357

\bibitem {Wang02b}
Q.A. Wang, Correlated electrons and generalized statistics, {\em Euro. Phys. J. B,}
{\bf31}(2003)75-79

\bibitem {Wang02c}
Q.A. Wang and A. Le M\'ehaut\'e, Extensive form of equilibrium nonextensive statistics,
{\em J. Math. Phys,} {\bf 43}(2002)5079-5089

\bibitem {Tsal94}
C. Tsallis, {\em Quimica Nova,\/} {\bf 17} (1994)468


\bibitem {Yama02}
Takuya Yamano, {\em Physica A,\/} {\bf 305}(2002)486

\bibitem {Agui03}
C. E. Aguiar and T. Kodama, {\em Physica A,\/} {\bf 320}(2003)371

\bibitem {Borges}
E.P. Borges, {\em J. Phys.A:Math.Gen.,\/} {\bf 31} (1998)5281-5288;

E.P. Borges, A possible deformed algebra and calculus inspired in nonextensive
thermostatistics, cond-mat/0304545


\bibitem {Kaniadakis}
G. Kaniadakis, {\em Phys. Rev. E,\/} {\bf 66}(2002)056125

G. Kaniadakis, {\em Physica A,\/} {\bf 296}(2001)405

\bibitem {Abe01}
S. Abe, {\em Phys. Rev. E,\/} {\bf 63}(2001)061105

\bibitem {Wang02d}
Q.A. Wang, L. Nivanen, A. Le M\'ehaut\'e and M. Pezeril, On the generalized entropy
pseudoadditivity for complex systems, {\em J. Phys. A,\/} {\bf 35}(2002)7003

\bibitem {Santos}
R.J.V. dos Santos, {\em J. Math. Phys.,\/} {\bf 38}(1997)4104

S. Abe, {\em Phys. Lett. A,\/} {\bf 271}(2000)74

\bibitem {Wang02e}
Q.A. Wang, {\em Phys. Lett. A,\/} {\bf 300}(2002)169

Q.A. Wang, {\em Chaos, Solitons $\&$ Fractals,\/} {\bf 14}(2002)765

\bibitem {Wang03}
Q.A. Wang, Incomplete information in fractal phase space, {\em Chaos, Solitons $\&$ Fractals,}
(2003), in press, cond-mat/0207647

Q.A. Wang, Measuring information growth in fractal phase space, cond-mat/0305540

\bibitem {Bear}
A.F. Beardon, An introduction to hyperbolic geometry, in Ergodic theory, symbolic
dynamics and hyperbolic spaces, Oxford University Press, New York, (1991)

\bibitem {Leme1}
A. Le M\'ehaut\'e, R. Nigmatullin, L. Nivanen, Fl\`eches du temps et g\'eom\'etrie
fractale, Hermes, Paris, (1998)

\bibitem {Leme2}
A. Le M\'ehaut\'e, L. Nivanen, Proceedings of SPIE, 4061(2000)180.


\end{thebibliography}
\end{document}